# MỐI TƯƠNG QUAN CỦA CÁC NHÂN TỐ ẢNH HƯỞNG TỚI VIỆC SỬ DỤNG ỨNG DỤNG BLUEZONE

Nguyễn Thế Vịnh[1*], Nguyễn Tuấn Anh[1], Nguyễn Hồng Tân[1], Lương Khắc Định[2]

[1]Khoa Công nghệ thông tin, Trường ĐH Công nghệ thông tin và Truyền thông, ĐH Thái Nguyên
[2]Khoa Công nghệ thông tin, Trường ĐH Hạ Long

* Email: vinhnt@ictu.edu.vn

## TÓM TẮT

Sự xuất hiện của đại dịch Covid-19 đã gây ra nhiều tác động tiêu cực đến mọi mặt của đời sống. Chính phủ đã áp dụng nhiều biện pháp để giảm thiểu sự ảnh hưởng và lây truyền của dịch bệnh. Trong số đó có việc áp dụng chuyển đổi số đối với việc quản lý và truy vết người bị nhiễm Covid thông qua phần mềm Bluezone (nay là PC-Covid). Tuy nhiên, việc cài đặt và sử dụng Bluezone lại không được như kỳ vọng. Vì vậy, nghiên cứu này tìm hiểu những nhân tố chính và sự ảnh hưởng của chúng tới ý định hành vi của người dùng về việc sử dụng phần mềm truy vết Bluezone. Phiếu khảo sát được gửi tới người dùng thông qua công cụ Google Form. Kết quả phân tích các nhân tố khám phá trên 224 đối tượng khảo sát cho thấy, có bốn nhân tố chính ảnh hưởng tới hành vi của người dùng, trong đó: sự tin tưởng và kỳ vọng hiệu quả, kỳ vọng nỗ lực, ảnh hưởng xã hội có tác động tích cực đến ý định hành vi của việc sử dụng phần mềm truy vết Bluezone; trong khi rủi ro về quyền riêng tư có ảnh hưởng tiêu cực đến hành vi này.

***Từ khóa:*** *EFA, SEM, UTAUT, tin tưởng, quyền riêng tư, Covid-19.*

## FACTORS INFLUENCING TO USE OF BLUEZONE

### ABSTRACT

The emergence of the Covid-19 pandemic has been causing many negative impacts on all aspects of life. The government has taken many measures to minimize the impact and transmission of the disease. Among them is the application of digital transformation to the management and tracing of people infected with Covid through the Bluezone app (now PC-Covid). However, using and installing Bluezone is not as expected. Therefore, this study aims to understand the main factors and their influence on the behavioral intention of users about using Bluezone. Surveys are sent to users through the Google Form tool. Experimental results through analysis of exploratory factors on 224 survey subjects show that there are 4 main factors affecting user behavior. Structural equation modeling indicates that trust, performance expectations, effort expectations, and social influence have a positive impact on behavioral intention of using Bluezone. Meanwhile, privacy risks have a negative effect on this behavior.

# 1. ĐẶT VẤN ĐỀ

Đại dịch Covid-19 xuất hiện vào cuối năm 2019 và bùng phát mạnh mẽ trong thời gian qua đã có những ảnh hưởng tiêu cực tới tất cả các quốc gia trên toàn thế giới (Whitelaw và c.s., 2020). Đứng trước vấn đề đó, chính phủ các quốc gia trên thế giới đã tiến hành nhiều biện pháp cấp bách nhằm hạn chế tầm ảnh hưởng, lây lan của dịch bệnh (Nguyen và c.s., 2021). Song song với các biện pháp tuyên truyền đến người dân về ý thức phòng chống dịch thông qua các phương tiện truyền thông, chính phủ Việt Nam cũng tiến hành nhiều biện pháp hỗ trợ nhằm truy vết tiếp xúc và cảnh báo người nhiễm Covid-19 (Le và c.s., 2021). Cụ thể, Bộ Y tế và Bộ Thông tin và Truyền thông đã phối hợp tạo ra ứng dụng Bluezone. Bluezone được coi là "cần thiết trong quá trình sinh hoạt hàng ngày, khi mọi người có tiếp xúc, ứng dụng trên điện thoại của họ sẽ tự *"nói chuyện"* với nhau" (baochinhphu.vn, 2020). Ứng dụng Bluezone được kỳ vọng là sẽ giúp ích cho các cơ quan nhà nước có thể nhanh chóng truy vết và quản lý được các ca nhiễm trong cộng đồng, người dân có thể nắm bắt được thông tin kịp thời để phòng dịch (Nguyen và c.s., 2021).

Mặc dù Bluezone được kỳ vọng sẽ mang lại hiệu quả tích cực cao và nhiều người sẽ sử dụng, nhưng số liệu thống kê thực tế lại không được như mong muốn (Nguyen và c.s., 2021). Tính đến 27 tháng 5 năm 2021, cả nước chỉ ghi nhận 33,48 triệu lượt tải (khoảng 34,7% so với tổng dân số), trong đó tập trung chủ yếu ở hai địa phương lớn là Hà Nội (3,1 triệu lượt cài đặt) và Thành phố Hồ Chí Minh (2,83 triệu lượt cài đặt). Ở chiều ngược lại, các tỉnh khác như Điện Biên, Kon Tum, Lai Châu, Bắc Kạn lại ghi nhận số lượng người tải ứng dụng Bluezone thấp nhất. Vì vậy, câu hỏi đặt ra là: Những yếu tố nào ảnh hưởng tới việc sử dụng phần mềm Bluezone?

Trả lời được câu hỏi nghiên cứu trên đóng vai trò quan trọng trong việc khuyến khích người dân tham gia, hỗ trợ phòng chống dịch trên môi trường số (Nguyen & Nguyen, 2022; Whitelaw và c.s., 2020). Có nhiều nghiên cứu trên thế giới tìm hiểu các yếu tố ảnh hưởng tới việc sử dụng phần mềm truy vết nói chung (Mbunge, 2020; Whitelaw và c.s., 2020), nhưng chưa có nghiên cứu nào được thực hiện ở Việt Nam trả lời cho câu hỏi trên một cách đầy đủ. Vì vậy nghiên cứu này có vị trí riêng biệt và cần thiết trong bối cảnh hiện nay, đặc biệt khi đại dịch Covid-19 vẫn chưa có dấu hiệu kết thúc do sự xuất hiện của các biến chủng mới. Nghiên cứu của Nguyen và c.s. (2021) mới chỉ dừng lại ở việc trích xuất được các nhân tố mà chưa xem xét đến mối tương quan giữa các nhân tố đó tới ý định sử dụng phần mềm Bluezone như thế nào. Chính vì vậy, nghiên cứu này được mở rộng bằng cách áp dụng mô hình phương trình cấu trúc nhằm đánh giá mối quan hệ giữa các yếu tố tới ý định sử dụng phần mềm Bluezone. Kết quả của bài báo được kỳ vọng sẽ có những đóng góp tích cực trong lĩnh vực nghiên cứu bao gồm: 1) việc khám phá ra các nhân tố chính ảnh hưởng tới ý định sử dụng phần mềm Bluezone, 2) đánh giá mối quan hệ giữa các yếu tố tới ý định sử dụng phần mềm Bluezone. Kết quả nghiên cứu sẽ là tài liệu tham khảo cho các nghiên cứu tương tự và là một trong các chỉ báo giúp các nhà quản lý điều chỉnh chính sách phù hợp nhằm nâng cao hiệu quả của ứng dụng truy vết.

# 2. MÔ HÌNH NGHIÊN CỨU VÀ CƠ SỞ LÝ THUYẾT

## 2.1. Tổng quan về mô hình nghiên cứu

Sự phát triển không ngừng của các thiết bị mới và phần mềm mới đã giúp cho người dùng trải nghiệm và giải quyết các vấn đề trong cuộc sống dễ dàng hơn. Tuy nhiên, không phải mọi công nghệ mới đều được người dùng chấp nhận và sử dụng. Để giảm thiểu các rủi ro trên, nhiều mô hình chấp nhận công nghệ được phát triển và áp dụng rộng rãi như: mô hình SOR – stimulus (kích thích), organism (chủ thể), response (phản hồi) – mô tả cách mà sinh vật, con người phản ứng, đáp lại với kích thích từ môi trường (Mehrabian & Russell, 1974), mô hình chấp nhận công nghệ – Technology Acceptance Model (TAM) (Davis, 1985), mô hình lý thuyết chấp nhận công nghệ hợp nhất (UTAUT). UTAUT được phát triển bằng việc kết hợp và tinh chỉnh tám mô hình trước đây thành một mô hình duy nhất để mô tả hành vi của người dùng với một hệ thống công nghệ thông tin (Venkatesh và c.s., 2003). Mô hình UTAUT chỉ ra có 4 yếu tố chính ảnh hưởng đến hành vi của người dùng bao gồm: kỳ vọng hiệu quả (performance expectancy), kì vọng nỗ lực (effort expectancy), ảnh hưởng xã hội (social influence), và các điều kiện thuận lợi (facilitating conditions). Ngoài ra còn có các yếu tố khác điều chỉnh đến ý định sử dụng như giới tính, độ tuổi, sự tự nguyện và kinh nghiệm. UTAUT được áp dụng rộng rãi trong nhiều lĩnh vực khác nhau (Jung và c.s., 2020, 2021; Nguyen, 2022). Trong nghiên cứu này, chúng tôi mở rộng mô hình UTAUT với hai nhân tố mới là sự riêng tư (privacy) và độ tin cậy (trust) được tham khảo từ những nghiên cứu tương tự (Arfi và c.s., 2021; Chopdar, 2022).

## 2.2. Cơ sở lý thuyết

*Kỳ vọng hiệu quả* (Performance Expectancy) được định nghĩa là mức độ mà một cá nhân tin rằng việc sử dụng hệ thống sẽ giúp họ đạt được hiệu quả trong công việc (Venkatesh và c.s., 2003). Năm yếu tố từ các mô hình khác nhau liên quan đến kỳ vọng hiệu quả là nhận thức phần mềm hữu ích, động lực bên ngoài, sự phù hợp với công việc, lợi thế tương đối và kỳ vọng kết quả.

*Kỳ vọng nỗ lực* (Effort Expectancy) được định nghĩa là mức độ dễ dàng liên quan đến việc sử dụng hệ thống (Venkatesh và c.s., 2003). Ba yếu tố từ các mô hình khác nhau liên quan đến kỳ vọng nỗ lực là nhận thức dễ sử dụng, độ phức tạp (mô hình sử dụng máy tính) và tính dễ dùng (mô hình khuếch tán đổi mới).

*Ảnh hưởng xã hội* (Social Influence) được định nghĩa là mức độ mà một cá nhân nhận thấy rằng những người khác quan trọng tin rằng họ nên sử dụng hệ thống mới (Venkatesh và c.s., 2003). Ba yếu tố từ các mô hình khác nhau liên quan đến ảnh hưởng xã hội là chuẩn chủ quan, yếu tố xã hội và hình ảnh.

*Các điều kiện thuận lợi* (Facilitating Conditions) được định nghĩa là "Mức độ mà một cá nhân tin rằng có sẵn cơ sở hạ tầng kỹ thuật và tổ chức để hỗ trợ việc sử dụng hệ thống" (Venkatesh và c.s., 2003). Venkatesh cho rằng các điều kiện thuận lợi không ảnh hưởng đến ý định hành vi, nhưng ảnh hưởng đến hành vi sử dụng. Các điều kiện thuận lợi liên quan đến sự sẵn có của nguồn lực và hỗ trợ cho các cá nhân sử dụng công nghệ.

*Rủi ro về quyền riêng tư* (Privacy Risk) được hiểu là mối quan ngại của người dùng về việc tiết lộ thông tin cá nhân (Arfi và c.s., 2021; Chopdar, 2022; Li, 2011). Nhiều nghiên cứu đã chỉ ra rằng rủi ro về quyền riêng tư có ảnh hưởng tới độ tin cậy của người dùng và gián tiếp ảnh hưởng đến ý định sử dụng hệ thống (Arfi và c.s., 2021; Bansal và c.s., 2010; Chopdar, 2022).

*Sự tin tưởng* (Trust) phản ánh sự sẵn sàng ở trong tình trạng dễ bị tổn thương dựa trên kỳ vọng tích cực đối với hành vi trong tương lai của yếu tố ngoại vi (Arfi và c.s., 2021; Chopdar, 2022). Nhiều nghiên cứu đã chỉ ra rằng sự tin tưởng có ảnh hưởng tới ý định hành vi và nhận thức rủi ro (Arfi và c.s., 2021; Chopdar, 2022).

## 3. PHƯƠNG PHÁP NGHIÊN CỨU

### 3.1. Đối tượng nghiên cứu

Phiếu khảo sát được tạo ra và gửi đến người dùng thông qua ứng dụng Zalo và mạng xã hội Facebook trong khoảng thời gian từ ngày 18/6/2021 đến ngày 21/6/2021. Số lượng ước lượng người dùng tham gia khảo sát là 400 người, tỷ lệ phản hồi là 73,75% (295 phản hồi), nhóm nghiên cứu loại bỏ 25 phản hồi do người dùng không cài đặt ứng dụng Bluezone, 41 câu trả lời không hợp lệ do chỉ chọn một lựa chọn duy nhất, 5 phản hồi không hoàn thành khảo sát. Tổng số dữ liệu cuối cùng để đưa vào phân tích là 224 (75,93%). Bảng 1 tổng hợp dữ liệu từ phiếu khảo sát, tỷ lệ nam chiếm 16,07%, trong khi đó tỷ lệ nữ chiếm 83,48%. Hơn một nửa đối tượng tham gia điều tra là sinh viên, học sinh trong độ tuổi từ 10 – 20 (52,68%), 27,23% nằm trong độ tuổi từ 21 – 30, 11,16% nằm trong độ tuổi 31 – 40%, số còn lại trên 41 tuổi chiếm 8,93%. Khu vực sinh sống của người dùng ứng dụng Bluezone chủ yếu tập trung ở khu vực thị xã, nông thôn và miền núi (52,23%), còn lại là ở các khu vực thành phố (28,57%) và quận/huyện (19,20%). Kết quả của phiếu khảo sát này cũng phù hợp với đặc tính vùng miền của tỉnh Thái Nguyên – là tỉnh miền núi.

## 3.2. Công cụ khảo sát

Sau khi nghiên cứu các câu hỏi dùng cho việc khảo sát dựa trên mô hình nghiên cứu (Arfi và c.s., 2021; Chopdar, 2022), 18 câu hỏi được nhóm tác giả lựa chọn và đưa vào nghiên cứu (xem

Bảng 2). Thang điểm Likert năm điểm (1 = Hoàn toàn không đồng ý, 2 = Không đồng ý, 3 = Trung lập, 4 = Đồng ý, 5 = Hoàn toàn đồng ý) được sử dụng cho mỗi câu hỏi.

**Bảng 1. Thông tin chung về đối tượng khảo sát**

| Thông tin chung | Số lượng | % |
|---|---|---|
| **Giới tính** | | |
| Nam | 36 | 16,07 |
| Nữ | 187 | 83,48 |
| Không xác định | 1 | 0,45 |
| **Độ tuổi** | | |
| 10 – 20 | 118 | 52,68 |
| 21 – 30 | 61 | 27,23 |
| 31 – 40 | 25 | 11,16 |
| Trên 40 tuổi | 20 | 8,93 |
| **Khu vực sinh sống** | | |
| Thành phố | 64 | 28,57 |
| Quận/huyện | 43 | 19,20 |
| Thị xã, nông thôn | 117 | 52,23 |
| **Tổng** | **224** | **100** |

## 3.3. Phân tích các nhân tố khám phá

Phân tích nhân tố khám phá (Explatory Factor Analysis - EFA) là một phương pháp thống kê dùng để rút gọn nhiều biến đo lường phụ thuộc lẫn nhau (đo được) thành một tập biến ít hơn (gọi là các nhân tố – không đo được trực tiếp) mà vẫn chứa đựng hầu hết nội dung thông tin của tập biến ban đầu (Hair Jr và c.s., 2009). EFA giả định rằng mỗi chỉ số trong một tập hợp các chỉ số là một hàm tuyến tính của một hoặc nhiều nhân tố chung và một nhân tố duy nhất. Các nhân tố chung là các yếu tố tiềm ẩn không thể quan sát được có ảnh hưởng đến nhiều hơn một chỉ số trong một tập hợp các chỉ số (Fabrigar & Wegener, 2012). Các nhân tố duy nhất là các biến tiềm ẩn được giả định chỉ ảnh hưởng đến một chỉ số từ một tập hợp các chỉ số và không tính đến mối tương quan giữa các chỉ số. Mục tiêu của mô hình nhân tố chung là tìm hiểu cấu trúc mối tương quan giữa các chỉ số bằng cách ước tính các mô hình mối quan hệ giữa các chỉ số và các nhân tố tiềm ẩn được lập chỉ mục gọi là tải nhân tố.

**Bảng 2. Bảng câu hỏi sử dụng khảo sát**

| Mã | Câu hỏi |
|---|---|
| *Kỳ vọng hiệu quả* (Venkatesh và c.s., 2003) | |
| PE1 | Sử dụng phần mềm Bluezone giúp tôi nắm bắt thông tin về Covid nhanh hơn. |
| PE2 | Sử dụng phần mềm Bluezone giúp tôi nâng cao hiệu quả về phòng tránh Covid. |
| PE3 | Sử dụng phần mềm Bluezone giúp tôi nắm bắt kịp thời các thông tin cần thiết nơi tôi sinh sống. |
| *Kỳ vọng nỗ lực* (Venkatesh và c.s., 2003) | |
| EE1 | Học cách sử dụng phần mềm Bluezone là tương đối dễ với tôi. |
| EE2 | Các chức năng và thao tác của Bluezone là rõ ràng và dễ hiểu. |
| EE3 | Phần mềm Bluezone là dễ sử dụng. |
| EE4 | Tôi dễ dàng sử dụng thành thạo phần mềm Bluezone. |
| *Ảnh hưởng xã hội* (Venkatesh và c.s., 2003) | |
| SI1 | Người thân trong gia đình tôi cho rằng tôi nên sử dụng phần mềm Bluezone. |
| SI2 | Bạn bè và đồng nghiệp tôi cho rằng tôi nên sử dụng phần mềm Bluezone. |
| SI3 | Tôi sử dụng phần mềm Bluezone là do được tuyên truyền từ các phương tiện truyền thông. |
| *Các điều kiện thuận lợi* (Venkatesh và c.s., 2003) | |
| FC1 | Tôi có thiết bị để cài đặt phần mềm Bluezone (ví dụ: điện thoại, máy tính bảng). |
| FC2 | Phần mềm Bluezone tương thích với các thiết bị của tôi. |
| FC3 | Tôi có sự hỗ trợ khi gặp trục trặc với phần mềm Bluezone. |
| *Rủi ro về quyền riêng tư* (Arfi và c.s., 2021; Chopdar, 2022) | |
| PR1 | Tôi nghĩ rằng việc sử dụng Bluezone sẽ khiến quyền riêng tư của tôi gặp rủi ro. |
| PR2 | Dữ liệu cá nhân của tôi có thể bị rò rỉ khi sử dụng phần mềm Bluezone. |
| *Sự tin tưởng (Trust)* (Arfi và c.s., 2021; Chopdar, 2022) | |
| T1 | Tôi tin rằng thông tin mà Bluezone cung cấp là đáng tin cậy. |
| T2 | Tôi tin tưởng việc sử dụng phần mềm Bluezone. |
| T3 | Bluezone cung cấp các chức năng mà người dùng cần. |

Nếu giá trị trung bình của một câu được tìm thấy là gần với 1 hoặc 5 thì nhóm nghiên cứu loại bỏ câu trả lời đó ra khỏi bảng số liệu vì nó có thể làm giảm tiêu chuẩn tương quan giữa các mục còn lại (J. J. Kim, 2011). Sau bước này, tính chuẩn mực trong phân phối đã được kiểm tra bằng cách kiểm tra độ lệch (skewness) và độ nhọn (kurtosis) trước khi tiến hành phân tích nhân tố khám phá. Vì tính chuẩn mực của phân phối đã được xác nhận, nên việc phân tích nhân tố khám phá được tiến hành thông qua việc sử dụng phần mềm SPSS 26 (Statistical Package for the Social Sciences).

Tiến trình phân tích nhân tố khám phá được bắt đầu bằng việc thu thập các giá trị riêng (eigenvalues) cho mỗi nhân tố. Tiếp theo, thang đo Kaiser-Meyer-Olkin (KMO) được sử dụng để đo về mức độ phù hợp của dữ liệu cho việc phân tích nhân tố (Goretzko và c.s., 2021). Giá trị của KMO thay đổi giữa 0 và 1 và các giá trị trên 0,5 thường được coi là đủ cho EFA (Goretzko và c.s., 2021; Schneeweiss & Mathes, 1995). Mức độ tương quan giữa các câu hỏi có đủ lớn để phân tích nhân tố có ý nghĩa thống kê hay không được kiểm tra thông qua phương pháp Bartlett. Chỉ khi kiểm định Bartlett có ý nghĩa thống kê (sig. < 0,05) thì các phân tích tiếp theo mới được tiến hành.

### 3.4. Mô hình phương trình cấu trúc

Sau khi có kết quả từ phân tích nhân tố khám phá, các nhân tố tìm được sẽ được sử dụng để tìm hiểu sự tác động của chúng đối với ý định hành vi của việc sử dụng phần mềm Bluezone. Mô hình phương trình cấu trúc (Structural Equation Modeling – SEM) được sử dụng để tìm hiểu sự tác động của các biến độc lập (nhân tố) đối với biến phụ thuộc (ý định hành vi) (Kline, 2015). SEM là một mô hình cấu trúc tuyến tính bao gồm các mô hình thống kê nhằm tìm lời giải thích mối quan hệ giữa các biến số (Kline, 2015). SEM được ứng dụng rộng rãi trong nhiều lĩnh vực với các tên gọi khác nhau như phân tích cấu trúc hiệp phương sai, phân tích biến ẩn, hoặc mô hình nhân quả. Mục đích của SEM là kiểm tra lý thuyết bằng cách chỉ định một mô hình đại diện cho các dự đoán của lý thuyết đó trong số các cấu trúc hợp lý được đo bằng các biến quan sát thích hợp.

## 4. KẾT QUẢ NGHIÊN CỨU

### 4.1. Phân tích nhân tố khám phá

EFA được thực hiện trên 18 câu hỏi với vòng quay Varimax. Kết quả phân tích từ phần mềm SPSS cho phép nhóm nghiên cứu trích xuất được giá trị đặc trưng cho từng nhân tố. Phép đo KMO đã xác minh tính thích hợp của việc lấy mẫu cho phép phân tích với giá trị là 0,889 (xem Bảng 3), cao hơn đề xuất của J. O. Kim & Mueller (1978) là 0,6.

**Bảng 3. Kiểm định KMO và Barlett**

| Kaiser-Meyer-Olkin | | 0,889 |
|---|---|---|
| Kiểm định Bartlett | Chi-Square | 2825,528 |
| | df | 153 |
| | Sig. | 0,000 |

Kiểm định Bartlett (Bartlett's test of sphericity) cho kết quả $\chi^2$ (153) = 2825,528, $\rho < 0,000$, chỉ ra rằng mối tương quan giữa các hạng mục câu hỏi là đủ lớn để tiến hành phân tích nhân tố khám phá.

Số liệu từ

Bảng 4 cho thấy có bốn nhân tố chính được hình thành từ tập 18 câu hỏi với giá trị đặc trưng lớn hơn 1. Nói cách khác, 18 câu hỏi này đóng góp 70,269% tầm quan trọng của các yếu tố tác động đến việc sử dụng ứng dụng Bluezone, 29,731% còn lại là do các yếu tố khác. Tỷ lệ phần trăm được giải thích theo từng nhân tố là: nhân tố 1 (46,749%), nhân tố 2 (10,563%), nhân tố 3 (6,587%) và nhân tố 4 (3,369%).

Dữ liệu trong Bảng 5 cho thấy có sự dịch chuyển về hạng mục câu hỏi giữa các nhân tố chính. Trong mô hình ban đầu, chúng tôi giả định rằng có sáu nhân tố chính ảnh hưởng tới việc sử dụng phần mềm Bluezone, tuy nhiên kết quả phân tích chỉ ra bốn nhân tố cơ bản phản ánh mối tương quan giữa các câu hỏi. Có một điểm đáng chú ý trong kết quả phân tích đó là nhóm nhân tố chính thứ hai và thứ tư vẫn giữ nguyên theo giả định ban đầu của nhóm tác giả, trong khi nhóm nhân tố chính thứ nhất được hình thành bằng việc kết hợp giữa hai yếu tố sự tin cậy (trust) và kỳ vọng hiệu quả (Performance Expectancy) – đặt lại tên là Hiệu quả tin cậy; Nhóm nhân tố thứ 3 được hình thành bằng việc kết hợp giữa ảnh hưởng xã hội và các điều kiện thuận lợi – đặt lại tên là Xã hội và Kỳ vọng hiệu quả. Hạng mục FC3 (Tôi có sự hỗ trợ khi gặp trục trặc với phần mềm Bluezone) bị loại bỏ sau quá trình phân tích.

Bảng 4. Các nhân tố chính

| Nhân tố | Giá trị đặc trưng khởi tạo | | | Tổng bình phương của hệ số tải nhân tố | | | Tổng bình phương của hệ số tải nhân tố xoay |
|---|---|---|---|---|---|---|---|
| | Tổng | % Phương sai | % Tích lũy | Tổng | % Phương sai | % Tích lũy | Tổng |
| 1 | 8,415 | 46,749 | 46,749 | 8,068 | 44,823 | 44,823 | 3,765 |
| 2 | 1,901 | 10,563 | 57,312 | 1,682 | 9,343 | 54,165 | 3,326 |
| 3 | 1,186 | 6,587 | 63,900 | 0,911 | 5,062 | 59,228 | 2,652 |
| 4 | 1,146 | 3,369 | 70,269 | 0,760 | 4,220 | 63,447 | 1,677 |
| 5 | 0,973 | 5,405 | 75,674 | | | | |
| 6 | 0,728 | 4,043 | 79,717 | | | | |

Bảng 5. Ma trận nhân tố xoay

| | 1 | 2 | 3 | 4 |
|---|---|---|---|---|
| T3 | 0,721 | | | |
| PE2 | 0,712 | | | |
| PE3 | 0,690 | | | |
| T2 | 0,649 | | | |
| PE1 | 0,575 | | | |
| T1 | 0,481 | | | |
| FC3 | | | | |
| EE1 | | 0,769 | | |
| EE2 | | 0,739 | | |
| EE3 | | 0,688 | | |
| EE4 | | 0,664 | | |
| SI2 | | | 0,796 | |
| SI1 | | | 0,671 | |
| FC1 | | | 0,614 | |
| FC2 | | | 0,566 | |
| SI3 | | | 0,375 | |
| PR1 | | | | 0,905 |
| PR2 | | | | 0,872 |

### 4.2. Mô hình phương trình cấu trúc

Dựa vào kết quả của phân tích nhân tố khám phá, nhóm nghiên cứu đưa ra các giả thiết sau:

*H1. Sự tin tưởng và kỳ vọng hiệu quả có ảnh hưởng tích cực tới ý định hành vi của việc sử dụng phần mềm Bluezone.*

*H2. Kỳ vọng nỗ lực có ảnh hưởng tích cực tới ý định hành vi của việc sử dụng phần mềm Bluezone.*

*H3. Ảnh hưởng xã hội có tác động tích cực tới ý định hành vi của việc sử dụng phần mềm Bluezone.*

*H4. Rủi ro về quyền riêng tư có ảnh hưởng tiêu cực tới ý định hành vi của việc sử dụng phần mềm Bluezone.*

Chỉ có các hạng mục có hệ số trong Bảng 5 lớn hơn 0,6 được giữ lại trong phân tích. Số mẫu tối thiểu cần thiết để phân tích có ý nghĩa thống kê theo công cụ tính toán của (Soper, 2022) là 166 (với 4 biến tiềm ẩn và 13 biến quan sát được). Số mẫu trong nghiên cứu là 224 lớn hơn so với số mẫu tối thiểu. Kỹ thuật phân tích thành phần có cấu trúc tổng quát (GSCA) được sử dụng để phân tích mô hình nghiên cứu được đề xuất do khả năng xử lý với kích thước mẫu nhỏ trong khi cần phân phối chuẩn nghiêm ngặt (Hwang & Takane, 2014). GSCA là một thành phần dựa trên mô hình phương trình cấu trúc có thể được sử dụng để mô phỏng các đường dẫn Bình phương tối thiểu một phần (PLS). Nghiên cứu này sử dụng phần mềm GSCA Pro trong việc ước lượng các tham số (Hwang và c.s., 2021).

Tính nhất quán của dữ liệu và các phép đo giá trị hội tụ cho mỗi nhân tố được thể hiện trong Bảng 6. Dillon-Goldstein's rho được sử dụng để đánh giá cho các yêu cầu về tính nhất quán và độ tin cậy bên trong của mỗi nhân tố (Hwang & Takane, 2014).

**Bảng 6. Độ tin cậy của thang đo**

| Nhân tố | Rho | AVE |
|---|---|---|
| Sự tin tưởng và kỳ vọng hiệu quả | 0,912 | 0,722 |
| Kỳ vọng nỗ lực | 0,940 | 0,796 |
| Ảnh hưởng xã hội | 0,897 | 0,746 |
| Rủi ro về quyền riêng tư | 0,947 | 0,899 |
| Ý định hành vi | 0,090 | 0,750 |

Hầu hết tất cả các giá trị, nằm trong khoảng từ 0,897 đến 0,947, đều lớn hơn 0,7, trên mức ước tính độ tin cậy có thể chấp nhận được (Hwang & Takane, 2014). Chúng tôi cũng đã xem xét giá trị phương sai trung bình được trích xuất (Average Variance Extracted – AVE) của mỗi biến tiềm ẩn để xác định xem biến có hội tụ hay không. Tất cả các giá trị AVE đều lớn hơn 0,5 (Hwang & Takane, 2014), nằm trong khoảng từ 0,722 đến 0,899, cho thấy độ tin cậy hội tụ.

**Bảng 7. Ước lượng hệ số tải (loadings)**

| | Ước lượng | SE | 95%CI_LB | 95%CI_UB |
|---|---|---|---|---|
| PE2 | 0,876 | 0,022 | 0,826 | 0,911 |
| PE3 | 0,850 | 0,031 | 0,786 | 0,904 |
| T2 | 0,833 | 0,031 | 0,782 | 0,893 |
| T3 | 0,839 | 0,027 | 0,763 | 0,887 |
| EE1 | 0,849 | 0,032 | 0,789 | 0,907 |
| EE2 | 0,912 | 0,017 | 0,873 | 0,939 |
| EE3 | 0,915 | 0,020 | 0,867 | 0,947 |
| EE4 | 0,890 | 0,019 | 0,851 | 0,932 |
| SI1 | 0,896 | 0,014 | 0,869 | 0,921 |
| SI2 | 0,943 | 0,008 | 0,924 | 0,955 |
| FC1 | 0,739 | 0,050 | 0,621 | 0,822 |
| PR1 | 0,949 | 0,008 | 0,934 | 0,968 |
| PR2 | 0,948 | 0,009 | 0,931 | 0,967 |
| BI1 | 0,893 | 0,029 | 0,836 | 0,939 |
| BI2 | 0,869 | 0,029 | 0,808 | 0,917 |
| BI3 | 0,835 | 0,043 | 0,716 | 0,889 |

Bảng 7 cho thấy hệ số tải của các hạng mục cùng với các tham số khác như sai số chuẩn (SE), khoảng tin cậy dưới (CI_LB) và khoảng tin cậy trên (CI_UB). Phương pháp Boostrap thực hiện với số mẫu lặp lại là 100 lần, giá trị trung bình của 100 lần lặp này được dùng để ước lượng giá trị gần đúng của tổng thể. Ở mức 0,05 alpha, ước tính tham số được coi là có ý nghĩa thống kê nếu 95% khoảng tin cậy không bao gồm giá trị 0. Kết quả Bảng 7 cho thấy tất cả các hạng mục đều đáng tin cậy và các ước lượng tải đều có ý nghĩa thống kê.

Kết quả phân tích từ phần mềm GSCA Pro cho các kết quả như: độ phù hợp của mô hình (Model FIT) là 0,59; độ phù hợp điều chỉnh của mô hình (Adjusted FIT - AFIT) là 0,586. Cả FIT và FIT điều chỉnh (AFIT) đều được sử dụng để điều tra sự khác biệt trong dữ liệu được giải thích bởi một cấu hình mô hình nhất định. Các giá trị FIT nằm trong khoảng từ 0 đến 1. Các đặc điểm và ý nghĩa của FIT và AFIT tương đương với $R^2$ và $R^2$ điều chỉnh trong hồi quy tuyến tính. Kết quả thực nghiệm của FIT và AFIT cho thấy mô hình lần lượt chiếm khoảng 59% và 58,6% tổng phương sai của tất cả các biến.

**Bảng 8. Ước tính hệ số đường dẫn**

| | Ước lượng | SE | 95%CI_LB | 95%CI_UB |
|---|---|---|---|---|
| Sự tin tưởng và kỳ vọng hiệu quả → Ý định hành vi sử dụng Bluezone (H1) | 0,218* | 0,105 | 0,029 | 0,049 |
| Kỳ vọng nỗ lực → Ý định hành vi sử dụng Bluezone (H2) | 0,116* | 0,084 | 0,05 | 0,290 |
| Ảnh hưởng xã hội → Ý định hành vi sử dụng Bluezone (H3) | 0,137* | 0,097 | 0,043 | 0,320 |
| Rủi ro về quyền riêng tư → Ý định hành vi sử dụng Bluezone (H4) | -0,06* | 0.63 | 0.06 | 0.185 |

*có ý nghĩa thống kê ở mức 0,05*

Bảng 8 trình bày các ước tính của hệ số đường dẫn của mô hình phương trình cấu trúc, cùng với sai số chuẩn và khoảng tin cậy 95% cận dưới và cận trên. Kết quả thực nghiệm cho thấy sự tin tưởng và kỳ vọng hiệu quả có ảnh hưởng tích cực tới ý định hành vi sử dụng phần mềm Bluezone (H1 = 0,218; SE = 0,105; CIs = 0,029 – 0,049). Tương tự, kỳ vọng nỗ lực có ảnh hưởng tích cực tới ý định hành vi (H2 = 0,016; SE = 0,084; CIs = 0,05 – 0,29). Ảnh hưởng xã hội cũng đóng góp vào ý định hành vi một cách tích cực (H3 = 0,137; SE = 0,097; CIs = 0,043 – 0,32) và cuối cùng rủi ro về quyền riêng tư có ảnh hưởng tiêu cực tới ý định hành vi sử dụng Bluezone của người dùng (H4 = -0,06; SE = 0,63; CIs = 0,06 – 0,185).

## 5. THẢO LUẬN

Đã hơn hai năm kể từ khi đại dịch Covid-19 xuất hiện, mặc dù số lượng ca nhiễm và tử vong đã giảm đáng kể so với giai đoạn đầu nhưng vẫn chưa có dấu hiệu nào cho thấy sự kết thúc của đại dịch này. Cùng với các biện pháp cách ly xã hội, tiêm vắc xin, khai báo trực tiếp bằng văn bản, việc ứng dụng công nghệ thông tin trong hỗ trợ đại dịch cũng đã và đang đem lại nhiều lợi ích nhất định. Hầu như các hoạt động xã hội đều đã được số hóa như họp trực tuyến, đặt hàng trực tuyến, thanh toán trực tuyến, giảng dạy trực tuyến... đến truy vết bằng công nghệ số. Các hoạt động này, cho dù không có đại dịch xảy ra, cũng là xu hướng tất yếu trong chuyển đổi số, nhưng sự xuất hiện của đại dịch khiến cho quá trình này được chuyển đổi nhanh hơn. Phần mềm Bluezone là sản phẩm kịp thời để ứng phó nhanh với đại dịch. Tuy nhiên, số lượng người dùng sử dụng liên tục lại không được như kỳ vọng. Điều đó dẫn đến ứng dụng công nghệ thông tin này chưa phát huy được hết sức mạnh. Do đó, ý thức về việc tích cực tham gia vào việc sử dụng ứng dụng Bluezone (hay PC-Covid) vẫn cần phải được nâng cao để giúp các nhà chức trách nhanh chóng tìm ra các giải pháp kịp thời.

Kết quả thực nghiệm từ mô hình phương trình cấu trúc cho thấy, cả bốn nhân tố thu được từ phân tích các nhân tố khám phá đều có ảnh hưởng tích cực hoặc tiêu cực tới ý định hành vi của người dùng đối với phần mềm Bluezone. Cụ thể, sự tin tưởng và kỳ vọng hiệu quả, kỳ vọng nỗ lực, ảnh hưởng xã hội có tác động tích cực đến ý định hành vi của việc sử dụng phần mềm truy vết Bluezone. Trong khi đó, rủi ro về quyền riêng tư có ảnh hưởng tiêu cực đến hành vi này.

Về mặt lý thuyết, kết quả của nghiên cứu này một lần nữa xác thực các mối quan hệ nguyên nhân – hậu quả đã được nghiên cứu và xác định ở trong mô hình phương trình cấu trúc, qua đó tạo thêm nhiều minh chứng cho sự tồn tại và ảnh hưởng của các nhân tố này. Những độc giả quan tâm hoặc các nhà nghiên cứu khác có thể tham khảo kết quả trên cho các nghiên cứu tương tự.

Về mặt thực tiễn, kết quả nghiên cứu là cơ sở để các nhà phát triển phần mềm, người quản lý đưa ra các chiến lược và giải pháp phù hợp để tăng cường ý định hành vi sử dụng phần mềm truy vết Bluezone. Cụ thể, đối với các nhân tố có ảnh hưởng tích cực, cần phải liên tục và cập nhật phần mềm sao cho nó thực sự mang lại hiệu quả hay nói cách khác, dữ liệu có được sử dụng tối ưu cho các nhà quản lý hay không. Hơn nữa, phần mềm phải nên thiết kế dễ sử dụng để bất kỳ ai cũng có thể tự thao tác. Ảnh hưởng xã hội cho thấy phương tiện truyền thông, gia đình, bạn bè và đồng nghiệp đóng vai trò quan trọng tới ý định hành vi, do đó việc tuyên truyền cũng nên tiếp tục được duy trì thông qua các phương tiện truyền thông khác nhau. Vì rủi ro về quyền riêng tư cũng đóng vai trò quyết định tới ý định, hành vi của người sử dụng, do đó các nhà quản lý, các nhà phát triển phần mềm, an ninh mạng cũng phải có các kỹ thuật, cơ chế, chính sách sử dụng và bảo vệ một cách phù hợp để giúp người dùng yên tâm hơn về dữ liệu cá nhân của mình.

Ngoài các yếu tố tích cực, nghiên cứu này cũng tồn tại một số giới hạn. Thứ nhất, việc lấy mẫu là không hoàn toàn ngẫu nhiên vì đối tượng tham gia nghiên cứu nằm trong mạng lưới của tác giả. Do đó việc khái quát hóa đến một số lượng người dùng lớn hơn cần phải được xem xét một cách kỹ lưỡng. Thứ hai, việc khảo sát chỉ được thực hiện trong một khoảng thời gian nhất định nên hành vi của đối tượng tham gia nghiên cứu có thể không nhất quán trong tương lai. Thứ ba, chỉ có một số các nhân tố được đưa vào phân tích trong mô hình phương trình cấu trúc, có thể tồn tại nhiều nhân tố khác cũng có tầm ảnh hưởng tới việc sử dụng Bluezone, do đó chúng tôi khuyến nghị các nhà nghiên cứu quan tâm tìm hiểu thêm các nhân tố mới này.

## 6. KẾT LUẬN

Nghiên cứu này khám phá các nhân tố và đánh giá sự ảnh hưởng của các nhân tố đó tới ý định hành vi của người dùng trong việc sử dụng phần mềm truy vết Bluezone. Mô hình lý thuyết thống nhất về chấp nhận và sử dụng công nghệ được mở rộng thêm hai nhân tố mới bao gồm sự tin tưởng và rủi ro về quyền riêng tư. Kết quả khảo sát từ 224 người dùng cho thấy có bốn nhân tố chính ảnh hưởng tới việc sử dụng phần mềm truy vết, trong đó có 3 nhân tố ảnh hưởng tích cực tới ý định hành vi, trong khi đó nhân tố rủi ro về quyền riêng tư có ảnh hưởng theo chiều ngược lại. Kết quả nghiên cứu đóng góp về mặt lý thuyết bằng cách giải thích sự ảnh hưởng của các nhân tố đối với ý định hành vi một cách tinh gọn hơn (giảm chiều từ 6 nhân tố xuống còn 4 nhân tố trong EFA) và xác thực các mối quan hệ nguyên nhân – hậu quả thông qua mô hình phương trình cấu trúc. Đồng thời, kết quả nghiên cứu cũng có thể được sử dụng trong thực tiễn giúp các nhà quản lý, nhà phát triển phần mềm, an ninh môi trường mạng có thêm cơ sở để tiếp tục hoàn thiện phần mềm truy vết Covid-19.

## LỜI CẢM ƠN

## TÀI LIỆU THAM KHẢO